\documentstyle[12pt,epsf,amsfonts]{article}
\newcommand{\be}{\begin{equation}}
\newcommand{\ee}{\end{equation}}
\newcommand{\bea}{\begin{eqnarray}}
\newcommand{\eea}{\end{eqnarray}}
\newcommand{\cU}{{\cal U}}

\newcommand{\CC}{{\mathbb C}}
\newcommand{\RR}{{\mathbb R}}

\newcommand{\ra}{\rightarrow}
\newcommand{\eps}{\epsilon}

\newcommand{\cO}{{\cal O}}
\newcommand{\cS}{{\cal S}}
\newcommand{\tS}{{\tilde S}}

\newcommand{\im}{{\rm Im}}
\newcommand{\sign}{{\rm sign}}
\newcommand{\stl}{\left|}
\newcommand{\str}{\right.}

\title{\vspace{-1in}\parbox{\linewidth}{\small\hfill
\shortstack{IASSNS-HEP-99/111}}
\vspace{0.6in}\\
\bf On The Universality Class Of Little String Theories}
  
\author{Anton Kapustin\thanks{email: kapustin@ias.edu}\\
{\small \it School of Natural Sciences, Institute for Advanced Study}\\
{\small \it Olden Lane, Princeton, NJ 08540}}

\begin{document}
\begin{titlepage}
\renewcommand{\thepage}{ }
\renewcommand{\today}{ }

\thispagestyle{empty}
\maketitle 

\begin{sloppypar} 
\begin{abstract} 
We propose that Little String Theories in six dimensions are
quasilocal quantum field theories. Such field theories obey
a modification of Wightman axioms which allows Wightman functions 
(i.e. vacuum expectation values of products of fundamental fields) to grow
exponentially in momentum space. Wightman functions of quasilocal fields in 
$x$-space violate microlocality at short distances. With additional assumptions
about the ultraviolet behavior of quasilocal fields, one can define 
approximately local observables 
associated to big enough compact regions. The minimum size of such a region
can be interpreted as the minimum distance which observables can probe.
We argue that for Little String Theories this distance is of order 
${\sqrt N}/M_s$.

\end{abstract} 
\end{sloppypar}
\end{titlepage}

\section{Introduction}

Poincar\'e-invariant theories in six dimensions have been much studied 
in the last few years. One of the reasons is that they describe 
$5+1$-dimensional branes which play an important role is string dualities.
In particular, the superconformal $(2,0)$ theory describing several
coincident M5-branes has attracted a lot of attention. Another reason is
that the very existence of consistent nontrivial Poincar\'e-invariant theories
in six dimensions came as a surprise. 

A standard strategy to construct
a nontrivial field theory is to take a free conformal theory and perturb
it by a relevant or marginally relevant operator other than the mass term. 
This method works well in dimension four or lower, but it is easy to
see that in higher dimensions free conformal theories do not have interesting
relevant or marginally relevant deformations. Hence until a few years ago it was 
believed that nontrivial quantum field theories do not exist in dimensions higher than 
four. 

$(2,0)$ and $(1,0)$ superconformal field theories describing various fivebranes
provided first examples to the contrary~\cite{Witone,Stro,GanHan,SeiWit,DLP}. 
Subsequently other related SCFTs in six dimensions have been 
discovered~\cite{Intri,BlumIntri}. 
A common feature of all these constructions is that they require taking a certain
limit in string theory or M-theory in which gravity and other bulk modes 
decouple from the fivebranes. In the simplest case, one takes $N$ coincident Type IIA 
fivebranes or $E_8\times E_8$ heterotic fivebranes and considers the limit $g_s\ra 0$, 
$M_s\ra\infty$. Here $g_s$ is the string coupling at spatial infinity, and 
$M_s=1/{\sqrt {\alpha'}}$ is the string scale. If gravity and the rest of
the bulk modes decouple, while the degrees of freedom living
on the brane remain interacting, one expects that that the brane degrees of
freedom are described in this limit by a nontrivial Poincar\'e-invariant
theory. In all known cases the argument for decoupling is indirect, and
the structure of the Poincar\'e-invariant theory is poorly understood.

Nevertheless, whenever one can argue decoupling in the limit $g_s\ra 0,
M_s\ra\infty$, it is believed that the Poincar\'e-invariant theory
is a local QFT; moreover, it is a (super)conformal QFT, since $M_s$,
the only scale in string theory, is taken to infinity.

For reasons mentioned above, these superconformal QFTs can not be obtained by
perturbing a free field theory with a local operator and therefore are not 
associated to any Lagrangian. But there are nonlocal theories in $5+1$ 
dimensions which flow to our SCFTs in the infrared: the so-called Little
String Theories (LSTs). Little String Theories first appeared in~\cite{BerRozSei}
where it was suggested that they describe M-theory compactified on $T^5$. 
The precise definition of LSTs as decoupled theories on fivebranes  
goes as follows~\cite{Sei}. One starts with $N$ coincident Type IIA or heterotic
fivebranes, as above, but now takes the limit $g_s\ra 0$ while keeping $M_s$ finite. 
Suppose this limit defines a Poincar\'e-invariant theory in $5+1$ dimensions.
Its infrared limit is equivalent to the limit $M_s\ra\infty$,
therefore by definition this theory flows to the SCFT of interest.
The difficult part is arguing that the decoupling really occurs.

All known nontrivial $(2,0)$ and $(1,0)$ superconformal theories in six 
dimension arise as the infrared limit of Little String Theories. In each case 
the parent Little String Theory has the same amount of supersymmetry, but does 
not have conformal or superconformal symmetry. Starting from a slightly different
brane configuration, one can also construct Little String Theories in 
six dimensions with $(1,1)$ supersymmetry~\cite{Sei}. According 
to Nahm's classification of superconformal algebras~\cite{Nahm}, such a theory 
cannot flow to a nontrivial superconformal theory in the infrared. Instead, at 
low energies $(1,1)$ theories reduce to $N=2$ $d=6$ super-Yang-Mills theories 
which are infrared-free.

The name ``Little String Theory'' has the following origin. Since we
did not send $M_s$ to infinity when taking the decoupling limit, it is 
natural to expect that the theory retains some stringy features. And indeed, 
one can argue that Little String Theories inherit from string theory such properties
as T-duality~\cite{Sei} and Hagedorn density of states~\cite{AB}.

The property of T-duality in particular seems to imply that Little
String Theories are not local quantum field theories (despite being
Poincar\'e-invariant). Intuition tells us that a quantum field theory always 
``knows'' on which pseudo-Riemannian manifold it lives, while T-duality means that
a Little String Theory on a torus of volume $V$ is indistinguishable from a
Little String Theory on a torus of volume $1/V$. (We make this argument more precise
in Section II.) 

Since a common lore says that the local quantum field theory framework is the 
only way to reconcile quantum mechanics, Poincar\'e-invariance, and macrocausality
(see e.g.~\cite{Wein}), we seem to be facing a puzzle.
As we explain below, the lore is incorrect if one is willing to sacrifice the 
notion of a strictly local observable. We will see that there is a 
way to modify the axioms of quantum field theory so that it is impossible
to construct observables whose support is a compact set. In these theories
local observables emerge only at distances much larger than a certain scale 
$\ell$. 

Nonlocal quantum field theories of this sort will be called quasilocal, 
since they violate causality only at distances shorter than $\ell$.
We conjecture that LSTs are quasilocal quantum field
theories with $\ell$ being of order ${\sqrt N}/M_s$. Here $N$ is the number of 
fivebranes.

Quasilocal field theory is quite an old subject (see \cite{FS} for a review). 
It was extensively studied in 60's and 70's in attempts to deal with nonrenormalizable
field theories. The main observation of the present paper is that 
the known properties of LSTs seem to fit perfectly into the framework of quasilocal 
field theory. 

The link between quasilocal field theories and LSTs is best seen 
from the holographic point of view~\cite{ABKS, PeetPolch}.
Holographic approach makes it clear that Little String Theories have properties
very similar to those of local QFTs. In particular, they have 
operators $\phi(p)$ which depend on the 6-momentum $p$ whose correlators ($=$ Wightman
functions) enjoy most of the usual properties. The only peculiarity of LSTs is 
that these correlators seem to grow exponentially in momentum 
space~\cite{PeetPolch, MinSei} (in local QFTs Wightman functions grow at most as 
a power). A related fact is the exponential growth of the density of states in 
LSTs~\cite{AB}. The importance of the exponential growth has been stressed by Aharony and 
Banks~\cite{AB}. These authors pointed out that exponential growth means that truly
local observables in LSTs do not exist. Rather, if the Wightman functions
grow as $\exp(\ell p)$, the minimal size the observables can probe
is of order $\ell$. Aharony and Banks suggested that the exponential growth
of Wightman functions in momentum space is the defining property of
LSTs.

In this paper we elaborate on the point made in~\cite{AB} and clarify the extent 
to which observables in LSTs can be localized.
We will see that the exponential growth of Green's functions
in momentum space is characteristic of a quasilocal field theory. Its $x$-space
counterpart is the fact that quasilocal fields
are very singular (but well-defined) distributions. 
The corresponding test functions are real-analytic and cannot have compact 
support. 

To appreciate the relation between the growth of distributions in momentum space 
and the properties of test functions in $x$-space, the following one-dimensional
example is helpful. Suppose we want to regard the functions 
${\tilde F}_\pm(p)=e^{\pm ap}$ of one 
real variable $p$ as distributions on a suitable space of test functions. (Here
$a$ is a positive real number). It is clear that the space of test functions
must include only functions which decay faster than $e^{-|a|p}$ at infinity. 
For example, one could take the space of smooth functions which are bounded by a 
multiple of $e^{-(|a|+\delta)p}$ for some positive $\delta$. What is the $x$-space 
analogue of this condition? The formal
Fourier transform of ${\tilde F}_\pm (p)$ is
\be
F_\pm(x)=\sum_{n=0}^\infty \frac{(\pm ia)^n}{n!} \delta^{(n)}(x).
\ee
Formally, the value of this distribution on a test function $f(x)$ is
\be\label{convser}
F_\pm(f)=\sum_{n=0}^\infty \frac{(\mp ia)^n}{n!} f^{(n)}(0).
\ee
For the distribution to be well-defined, this sum must converge. This means that
the derivatives of $f$ must not grow too fast. In fact, if the 
series (\ref{convser}) is convergent, then the Taylor expansion
for $f(x)$ around the point $x=0$ is convergent for $|x|<a$. Similarly, if we 
consider momentum-space
distributions of the form $e^{ibp\pm a p}$ and require that their Fourier transform
be well-defined, we will find that the Taylor expansion for $f(x)$ must converge
for $|x-b|<a$. Hence, if $b$ is allowed to be arbitrary, then the test functions
in $x$-space must be real-analytic. Moreover, they can be continued 
analytically off the real axis to a strip of width $a$.

The above arguments can be extended to the general $d$-dimensional case. 
We will see that if distributions
grow exponentially in momentum space, the test functions in 
$x$-space must be real-analytic. An operator
smeared with such a test function is not a local observable.
Furthermore, the microlocality condition does not make any sense for such
fields. Indeed, it says that if $f$ and $g$ are test functions whose supports 
are causally disconnected, then $\phi(f)$ and $\phi(g)$ commute (or anticommute).
This condition is empty for quasilocal theories,
since there are no test functions whose supports are causally disconnected!

One can nevertheless formulate a version of the microlocality axiom which does make
sense. To formulate this axiom, we first need to define
approximately local (AL) observables. The basic idea is to consider a
sequence of test functions $\{f_\nu\}$ which get more and more concentrated
on a certain compact set $M$. (The precise definition of what ``concentrated''
means will be given later.) Then we say that observables $\phi(f_\nu)$ 
are approximately local, and $M$ is their quasisupport. Thus quasisupport is
an attribute of a sequence of observables rather than of a single observable.
The quasilocality axiom says that AL observables whose quasisupports are space-like separated
approximately commute (or anticommute). This means that if $M$ and $N$
are space-like separated, and $\phi(f_\nu)$ and $\phi(g_\nu)$ are sequences of
AL observables with quasisupports $M$ and $N$, respectively, then 
\be
[\phi(f_\nu),\phi(g_\nu)]_-\ra 0\ or \ [\phi(f_\nu),\phi(g_\nu)]_+\ as\ \nu\ra\infty. 
\ee
The quasilocality axiom ensures locality ``in the large.'' Nonlocal QFTs which
satisfy this axiom are called quasilocal. Our proposal is that LSTs are quasilocal
field theories.

The outline of the paper is as follows. In Section II we summarize the known
properties of Little String Theories and argue that they cannot be
local QFTs. In Section III we explain how the axioms of local quantum
field theory should be modified
in order to incorporate Wightman functions which grow exponentially
in momentum space and why this leads to violations of locality at short 
distances. In Section IV we discuss how to define approximately local observables
in nonlocal theories. It turns out that nontrivial observables approximately localized
on a compact set $M$ can be defined for a special class of such theories, and only if 
$M$ is big enough. The precise definition of ``big enough'' sets 
depends on the reference frame. The minimal size of the support of an observable 
sets the smallest distance which observables in a quasilocal theory can probe. 
We argue that for Little String
Theories this distance is of order ${\sqrt N}/M_s$. In Section V we suggest 
directions for future work.

\section{Properties of Little String Theories}

Let us summarize what is known about LSTs in general.

1. LSTs are quantum-mechanical theories. This means that a state in LST is a ray 
in a Hilbert space $V$, and observables are self-adjoint linear operators on $V$.
Even in ordinary quantum field theory many important observables (energy, for example)
are unbounded operators which are only defined on a dense subset of $V$. Presumably the 
same is true about observables in LSTs.

2. LSTs are Poincar\'e-invariant, i.e. there is a unitary representation 
of the Poincar\'e group acting on $V$. In particular, there is a Hamiltonian
(the generator of time translations) which is an unbounded self-adjoint
operator on a dense subset of $V$. Furthermore, this operator has nonnegative spectrum.
For all known LSTs this holds because they are supersymmetric.

3. Among observables of LSTs there are operators $\phi_R(p)$ labeled by an 
irreducible finite-dimensional representation $R$ of $Spin(1,5)$ and the ``momentum''
$p\in \RR^{1,5}$~\cite{ABKS}. These operators can be thought of as functions on $\RR^{1,5}$
valued in the tensor product $End(V)\otimes R$. They are covariant with respect to 
the Poincar\'e group, i.e. for any element $(a,\Lambda), a\in\RR^{1,5}, 
\Lambda\in SO(1,5)$, of the Poincar\'e group we have
\be
U(a,\Lambda)\phi_R(p)U(a,\Lambda)^{-1}=e^{ipa} R(\Lambda)\phi(\Lambda^{-1}p).
\ee

4. The Hilbert space $V$ has a distinguished state $\Omega$ (vacuum) which is
Poincar\'e-invariant. Vacuum expectation values 
\be
\left(\Omega, \phi_{R_1}(p)\ldots \phi_{R_n}(p_n)\Omega\right) \sim 
\delta^6(p_1+\ldots +p_n) 
{\tilde W}_n(p_1-p_2,\ldots, p_{n-1}-p_n)
\ee
appear to grow exponentially for large momenta. For example, the 2-point
function seems to be growing as~\cite{PeetPolch, MinSei}
\be\label{growth}
{\tilde W}_2(p)\sim exp\left(\frac{c M_s p}{\sqrt N}\right),
\ee
where $N$ is the number of 5-branes, and $c$ is a numerical constant of order 
$1$.
(The authors of \cite{MinSei} chose to remove this exponential growth by making
a multiplicative renormalization of operators in momentum space. However, it appears
that the growth is common to all correlators and is related to the growth of the
density of states, see Property 7. Multiplicative renormalization of fields 
is not sufficient to make higher-order correlators polynomially bounded, except
in the case of a Gaussian theory.)

Strictly speaking, Eq.~(\ref{growth}) has been established only for large $N$ and 
for momenta in the range $M_s/{\sqrt N} \ll p\ll M_s$. However, it is plausible that 
this growth continues for $p\gg M_s$ (see Property 7 below).

5. Operators in LSTs obey the usual spin-statistics relation.

6. In the infrared LSTs flow to local quantum field theories.

7. The density of states of an LST grows exponentially at large energies~\cite{AB}.
Equivalently, the entropy per unit volume of the microcanonical ensemble is
\be
s\sim \frac{\eps {\sqrt N}}{M_s}
\ee
if the energy density $\eps$ is large, $\eps\gg M_s^6$. Consequently, the canonical 
ensemble is defined only for temperatures $T<M_s/{\sqrt N}$.

8. An LST on a manifold of the form $\RR^{1,5-n}\times T^n$ where $T^n$ is an 
$n$-dimensional torus with a flat metric is equivalent to a (in
general different) LST with the same $M_s$ on a manifold $\RR^{1,5-n}\times\hat{T}^n$~\cite{Sei}.
Here $\hat{T}^n$ is the dual torus. This means that different Little String 
Theories are related by T-dualities when compactified on tori. For example, the 
LST of parallel Type IIA 5-branes is mapped by a T-duality to the LST of 
parallel Type IIA or IIB 5-branes, depending on whether
$n$ is even or odd.

9. Some further properties of LSTs are discussed in~\cite{AhBer,GKP,GK}.

Property 8 is particularly striking. One's first reaction is that
a local quantum field theory cannot have a T-duality, and that only a string theory
of some kind would fit the bill. 

The former claim can be argued 
as follows. A local quantum field theory has local observables associated to compact
sets. According to the microlocality axiom, these observables commute at space-like separations. 
Thus by looking at the structure of the algebra of observables one can reconstruct unambiguously 
the causal structure of space-time, i.e. the position of light-cones. In other words, one
can reconstruct the conformal structure of space-time. Since in general a flat torus
and its dual are not conformally equivalent, a local quantum field theory cannot have
T-duality.

The claim that only a string theory can enjoy T-duality also sounds plausible. However
since the only string theory we know of is critical string theory, and there is
no agreement on how to define ``string theory in general'', this claim is almost
devoid of content. We will argue below that Little String Theories are a kind of 
quasilocal field theories which do not have truly local observables.

\section{Little String Theories as Quasilocal Field Theories}

\subsection{The Space of Test Functions}

In local quantum field theory fields are operator-valued
distributions~\cite{PCT}. This means that although the value of a field
at a point is not a well-defined observable, the field smeared with
a test function $f\in \cS$ is well-defined. Usually $\cS$ is taken to be
the space of infinitely differentiable functions which decay at infinity
faster than any negative power (the Schwartz space). The corresponding
distributions (i.e. continuous linear functionals in the standard topology on $\cS$) 
are called tempered distributions. Thus in local quantum field theory local fields
are operator-valued tempered distributions. 

The choice of the space of test functions seems like a technicality, but in
fact it has important physical consequences. For example, since the Fourier
transform of a function $f\in \cS$ is again an element of $\cS$~\cite{GelfShi}, tempered
distributions can grow at most as a polynomial in momentum space. This implies
that the correlators of local operators can grow at most as a power of momentum. 
If we want 
to accommodate operators whose matrix elements grow faster than a polynomial, one has
to work with more singular distributions, which are defined on a smaller space of test 
functions.

In view of properties 4 and 7, it seems very likely that Wightman functions (i.e.
vacuum expectation values) of operators $\phi_R(p)$ in Little String Theories
grow exponentially with momenta. More precisely, by positivity of energy the function 
${\tilde W}_n(q_1,\ldots, q_{n-1})$ vanishes when any of its arguments is outside
the forward light-cone $V_+$, but inside $V_+^{n-1}$ it appears to be bounded by 
a multiple of
\be\label{Whighen}
\exp\left(+\ell(|q_1|+\ldots |q_{n-1}|\right)),
\ee
where $|q|=\sqrt {q^2}$, and $\ell$ is of order ${\sqrt N}/M_s$. 
Clearly, such functions are not tempered distributions.\footnote{It is tempting to
call them ill-tempered distributions, but probably the name ``distributions of
exponential growth'' is more suitable.} Our first task is to find the
right space of test functions which could be used to smear ${\tilde W}_n$.

A necessary requirement on the test functions is that they be infinitely 
differentiable and decayed exponentially fast (in momentum space). The former 
requirement is necessary if we want the product of the field operator $\phi_R(x)$
and a polynomial of $x$ to be well-defined. The latter requirement comes from 
the exponential growth of the Wightman 
functions in momentum space. We also want the space of test functions to be Lorenz-invariant.
Finally, the space of test functions should
to be sufficiently ``nice''. As a minimum, we want it to be a complete countably normed
space in which the nuclear theorem holds, see e.g.~\cite{PCT}.

A convenient class of spaces of test functions was defined by A.~Jaffe~\cite{Jaffe}. Given
a function $g(t), t\in \RR$, such that $g(t^2)$ is entire~\footnote{By an entire function
on $\RR^m$ we mean a real-analytic function whose Taylor series has an infinite radius of 
convergence. An entire function can be analytically continued to a holomorphic function on 
$\CC^m$.} and positive, Jaffe defines
a space $\tS_g$ which consists of all functions on $\RR^d$ which are infinitely-differentiable
and for which all the norms 
\be\label{norms}
||f(p)||_n=\sup_{p;|m|\leq n} g(n||p||^2)|D^mf(p)|, \quad n=1,2,\ldots,
\ee
are finite. Here $m=(m_1,\ldots,m_d)$ is a polyindex, $|m|=\sum_i m_i,$ and
$||p||$ is an arbitrary Euclidean norm on $\RR^d$. Loosely speaking, the finiteness of the 
norms (\ref{norms}) means that a test functions and all its derivatives decay at 
infinity faster than $1/g(n||p||^2)$ for any positive $n$.
Despite appearances, $\tS_g$ does not depend
on the choice of the Euclidean norm, but it does depend on the rate of growth of $g(t^2)$
at infinity. If we define convergence on $\tS_g$ 
using the family of norms~(\ref{norms}), it becomes a complete countably normed space. 
It is easy to see that $\tS_g$ is Lorenz-invariant. In addition,
if $g$ satisfies 
\be
\frac{g(n||p||^2)}{g(n'||p||^2)}\ \emph{is an integrable function for all $n$ and 
sufficiently
large $n'$},
\ee
the nuclear theorem holds~\cite{GelfShi}.

Let us denote the Fourier transform of $\tS_g$ by $S_g$. We think of $\tS_g$ as the space
of test functions in momentum space, so $S_g$ is the space of test-functions in
$x$-space. The spaces $S_g$ can be used to define quantum field theories whose ultraviolet 
behavior is more singular than that of Wightman QFTs. Their localizability properties
depend on the rate of growth of $g$ at infinity.

A.~Jaffe showed~\cite{Jaffe} that if the function $g$ satisfies
\be\label{Ostr}
\int_0^\infty \frac{\log g(t^2)}{1+t^2} dt < \infty,
\ee
then the Fourier transform of $\tS_g$ contains many functions with compact support, so
that the microlocality axiom can be formulated in the usual manner.
QFTs based on the space $S_g$ with $g$ satisfying (\ref{Ostr}) 
are called strictly localizable~\cite{Jaffe}. Such QFTs have properties which 
are not very different from the properties of Wightman QFTs. 

Conversely, if $g$ does not satisfy (\ref{Ostr}), 
there are no functions with compact support among the test functions in $x$-space 
(except identical zero). For Little String Theories 
we want to take
\be
g(t)=e^{\sqrt t}.
\ee
This choice of $g$ is dictated by the exponential growth of the Wightman functions
~(\ref{Whighen}). With this choice of $g$ the test functions in momentum space decrease 
faster than any linear exponential, thus ensuring that the value of the Wightman 
functional is well-defined on any 
$n$-tuple $$(f_1,\ldots,f_n),\ f_1,\ldots,f_n\in \tS_g.$$
Since the condition~(\ref{Ostr}) is violated, the test functions in $x$-space
cannot have compact support. In fact, one can show that the test functions in 
$x$-space are entire. An entire function which vanishes in an open set is identically zero;
in particular the support of any nontrivial entire function is the whole 
$\RR^d$.

\subsection{Analytic properties of Wightman functions in LSTs} 

In this section we discuss the analyticity properties of Wightman functions in LSTs, 
assuming that LSTs are quantum field theories based on the Jaffe space $S_g$ with 
$g(t)=\exp(\sqrt t)$. 
Our motivation is the following. Recall that in local QFTs Wightman functions obey certain
symmetry properties as a consequence of microlocality, and conversely microlocality
follows from these symmetry properties~\cite{PCT,BLT}. One possible way to 
ensure locality of LSTs ``in the large'' is to impose a similar symmetry requirement
on their Wightman functions. However, before we do this, we need to understand
at which points the values of Wightman functions are well-defined. After all, 
``Wightman functions'' are really distributions, and pretty singular ones at that.
Luckily, this problem was studied in detail in the literature on nonlocal field 
theories. Below we summarize some of this work following~\cite{FI}.

In a local quantum field theory the microlocality axiom implies the permutation 
symmetry of Wightman functions,
\be\label{symm}
W_n(x_1,\ldots,x_i,x_{i+1},\ldots,x_n)=W_n(x_1,\ldots,x_{i+1},x_i,\ldots,x_n),
\ee
for $(x_i-x_{i+1})^2<0$. (To simplify our discussion, we will restrict ourselves
to bosonic fields here.) 
Since ``Wightman functions'' are not really functions, their values at points are not 
well-defined in general, and the meaning of Eq.~(\ref{symm}) must be clarified.

One way to interpret Eq.~(\ref{symm}) is to smear it with a product of test functions
$$f_1(x_1)\ldots f_n(x_n)$$ such that the supports of $f_i$ and $f_{i+1}$ are
space-like separated. However this approach does not extend to QFTs with
entire test functions. A more sophisticated interpretation of Eq.~(\ref{symm})
makes use of the fact that at certain special points the values of Wightman 
functionals are well-defined.
Recall that Wightman functionals can be regarded as boundary values
of certain holomorphic functions~\cite{PCT}. 
In other words, one can define ``analytic 
continuation'' of the $n$-point Wightman functional to complex values of its arguments,
and this continuation is a holomorphic function in a certain open set ${\cal T}_n\subset
\CC^{dn}$. ${\cal T}_n$ is called the forward tube and is given by
\be
{\cal T}_n=\left\{x+iy\stl x,y\in \RR^{dn};\zeta_i(y)\in V_+, i=1,\ldots,n-1\str\right\}.
\ee
Here $V_+$ is the forward light-cone in $\RR^d$, and for any point 
$(x_1,\ldots,x_n)\in\RR^{dn}$ we denote $\zeta_i(x)=x_i-x_{i+1}, i=1,\ldots,n-1$.
The real points are on the boundary of ${\cal T}_n$. The possibility
of ``analytic continuation'' is a consequence positivity of energy in local quantum
field theory, see \cite{PCT} for details. (Really, ``analytic continuation'' 
is simply the Laplace transform of momentum space Wightman functions which is 
well-defined because momentum space Wightman functions vanish outside the forward 
light-cone.)  If we apply complexified Lorenz 
transformations to the forward tube ${\cal T}_n$, we get the so-called extended forward 
tube ${\cal T}_n^{ext}$. By 
Lorenz-invariance, the ``analytic continuations'' of Wightman functionals
are holomorphic in the extended forward tube~\cite{PCT}. Now it is crucial that 
${\cal T}_n^{ext}$ includes real points (usually called Jost points).  
Jost points form an open set which we will call the Jost domain. 
Wightman functionals can be regarded as ordinary functions at all Jost points. 
When both $$(x_1,\ldots,x_i,x_{i+1},\ldots,x_n)$$ and
$$(x_1,\ldots,x_{i+1},x_i,\ldots,x_n)$$ belong to the Jost
domain, Eq.~(\ref{symm}) admits a straightforward interpretation.

Which real points belong to the Jost domain? The answer it simple for a 2-point 
function: $(x_1,x_2)$ is a Jost point if and only if $(x_1-x_2)^2<0$. Thus the 2-point
Wightman function is analytic for space-like separated points. Eq.~(\ref{symm})
simply says that $W_2$ is a symmetric function of its arguments when its arguments
are space-like separated. The situation for the
$n$-point functions is similar, except that the shape of the Jost domain is somewhat 
more complicated~\cite{PCT}: it is not sufficient to require that all vectors 
$x_i-x_j$ be space-like, one should also require that every point in the convex hull 
of the points $\zeta_i(x)=x_i-x_{i+1}, i=0,\ldots,n-1$ be space-like. This condition is 
also a necessary one for $(x_1,\ldots,x_n)$ to belong to the Jost domain. 

One can prove~(see e.g.~\cite{BLT}) that the permutation symmetry of Wightman 
functionals in the Jost domain is equivalent to microlocality, if the Wightman 
functionals are assumed to be tempered.

After this brief review of the analytic properties of Wightman functions in
a local QFT, let us return to QFTs based on the space $S_g$ with $g(t)=
\exp(\sqrt t)$. The analytic properties of Wightman functions in these theories
were described in~\cite{FI}. For a given Wightman function one can define
the analogue of the forward tube $\cU_n$ as the region in $\CC^{dn}$ where the 
Laplace transform of the momentum-space Wightman function is well-defined. It 
turns out that this domain is given by
\be
\cU_n=\left\{x+iy\stl x\in \RR^{dn}; y\in \bigcup_\eta V_\ell^n(\eta)\str\right\}.
\ee
Here $\ell$ is a positive real number, $\eta\in \RR^d$ is a unit time-like 
vector, and $V_\ell^n(\eta)$ is a domain in $\RR^{dn}$ given by
\be
V_\ell^n(\eta)=\left\{(x_1,\ldots,x_n)\in \RR^{dn}\stl
\zeta_i(x)\cdot\eta-\sqrt{(\zeta_i(x)\cdot \eta)^2-\zeta_i(x)^2}>\ell, \forall i
\str\right\}.
\ee
The meaning of the above equation is easiest to see in the frame where 
$\eta=(1,\vec{0})$. Then $V_\ell^n(\eta)$ becomes
\be
V_\ell^n(\eta)=\left\{(x_1,\ldots,x_n)\in \RR^{dn}\stl
\zeta_{i0}(x)-|\vec{\zeta}_i(x)|>\ell, \forall i\str\right\}.
\ee
It is easy to see that for $\ell>0$ the domain $\cU_n$ is a proper
subset of ${\cal T}_n$, and in the limit $\ell\ra 0$ the two domains coincide.

The parameter $\ell$ can be different for different 
Wightman functions and characterizes the nonlocality
of a given Wightman function. For $\ell>0$ the closure of $\cU_n$ does not 
include real points, unlike in a local QFT. Thus Wightman functions in $x$-space 
are not boundary values of holomorphic functions, in general. 

As before, Lorenz invariance implies that the Wightman functions are analytic
in the domain $\cU_n^{ext}$ obtained by applying complexified Lorenz
transformations to $\cU_n$. The Jost domain is defined as the set of real points
of $\cU_n^{ext}$. 

It is of interest to determine the precise shape of the Jost domain in the nonlocal
case. This has been done in~\cite{FI}. It is rather obvious that for
$\ell>0$ the ``nonlocal'' Jost domain is a proper subset of the ``local'' Jost domain.
For $n=2$ the ``nonlocal'' Jost domain is given by
\be
(x_1-x_2)^2<-\ell^2.
\ee
For $n>2$ a natural guess for what a ``nonlocal'' Jost domain is the following.
Let $J_n^\ell$ be the set of all points
$x\in \RR^{nd}$ such that the convex hull of $\zeta_1(x),\ldots,\zeta_{n-1}(x)$ 
belongs to the hyperboloid $\zeta^2<-\ell^2$. For $\ell\ra 0$ $J_n^\ell$ reduces
to the ``local'' Jost domain, and is a natural candidate to be the ``nonlocal''
Jost domain. This guess is incorrect. The actual shape of the ``nonlocal'' Jost 
domain
is rather more complicated, see~\cite{FI} for details. For our purposes it is
sufficient to know that the ``nonlocal'' Jost domain is a proper
subset of $J_n^\ell$~\cite{FI}. Since the closure of $J_n^\ell$ does not
contain any light-like separated points, the same is true about the ``nonlocal''
Jost domain. This means that in nonlocal field theories Wightman functions
may have singularities outside the light-cone. We will see some examples of this
below.

In view of the above, there is an obvious reformulation of the microlocality 
axiom which also makes sense for nonlocal field theories based on the space $S_g$. 
We simply require (anti)symmetry of Wightman functions
in their ``nonlocal'' Jost domains.
We will call this symmetry requirement weak quasilocality (strong quasilocality
will be discussed in the next section). Nonlocal field theories whose Wightman 
functions satisfy the weak quasilocality axiom will be called weakly quasilocal. 

In a general weakly quasilocal field theory $\ell$ can grow without bound as the 
order of the Wightman function
increases. Then the theory is nonlocal at all scales, if one studies a 
sufficiently complicated Wightman function. In Little String Theories the
exponential growth of Wightman functions is caused by the exponential growth of the
density of states. Thus it is natural to suspect that in LSTs $\ell$ is bounded from 
above by a quantity of order ${\sqrt N}/M_s$. In other words, LSTs are probably
``more local'' than a generic weakly quasilocal field theory. In Section IV
we will argue that LSTs satisfy a stronger condition 
(strong quasilocality) which implies weak quasilocality, boundedness of $\ell$,
as well as the existence of approximately local observables.

\subsection{A sample 2-point function}\label{toy}

Let us illustrate the preceding discussion with a simple example of a 2-point function
of a real scalar field. 
Poincar\'e-invariance and positivity of energy imply that the most general
2-point function in momentum space has the form
\be\label{twopoint}
{\tilde W}_2(p)=\theta(p_0)\theta(p^2) \sigma(p^2),
\ee
where the function $\sigma$ is positive and measurable. $\sigma(t)$ is the density of
states times some formfactor. In a local QFT $\sigma$ cannot 
grow faster than a polynomial, but in a weakly quasilocal QFT we only require 
that $\sigma(t)$ be bounded by $\exp(\ell{\sqrt t})$ for some $\ell$. In particular, 
we know that in LSTs the density of states grows like $\exp(\ell E)$ with $\ell$ of 
order ${\sqrt N}/M_s$. If the formfactor does not decrease exponentially, the function 
$\sigma(t)$ will grow like $\exp(\ell\sqrt t)$. Computation in \cite{MinSei,PeetPolch} seem to
support this conjecture.

As mentioned above, the Jost domain for a 2-point function is
\be\label{condspace}
(x_1-x_2)^2<-\ell^2.
\ee
The weak quasilocality condition is satisfied automatically, because $W_2(x)$ is
a function of $x^2$ by Lorenz invariance.

Given that the Wightman functional is a symmetric function of its arguments in the
region (\ref{condspace}) essentially
by virtue of Lorenz invariance, one may ask how it can fail to be symmetric
for all space-like points. The answer is very simple. In a quasilocal theory 
the Wightman function may have poles in the region $-\ell^2\leq (x_1-x_2)^2<0$. Thus one
needs a prescription how to treat these poles. The correct prescription involves
factors like $\sign(t_1-t_2)$, where $t_1,t_2$ are the time-like components of $x_1,x_2$.
Such factors are Lorenz-invariant, but not symmetric under the exchange of $x_1$ and 
$x_2$.

To illustrate this point, let us consider the following simple example in $d=4$
borrowed from~\cite{FI}.
Set 
\be\label{sigma}
\sigma=e^{\ell\sqrt t}/{\sqrt t}.
\ee
The Fourier integral 
\be
W_2(z)=\int d^4 p e^{-ipz} {\tilde W}_2(p)
\ee
converges in the region $(\im\  z)^2>\ell^2.$ In this region we get
\be\label{explicit}
W_2(z)=\frac{1}{2\pi^2(z^2+\ell^2)}\left\{\frac{2\ell}{z^2}+\frac{1}{\sqrt{z^2+\ell^2}}
\log \frac{\ell+\sqrt{\ell^2+z^2}}{\ell-\sqrt{\ell^2+z^2}}\right\}.
\ee
If we formally continue $W_2(z)$ to real values of $z$, we find a pole at 
$z^2=-\ell^2$ and a branch point at $z^2=0$. The branch point at $z^2=0$ is usually 
present in a local quantum field theory as well, but the pole at space-like $z$ is 
possible only in a nonlocal theory. Because of this pole, the integral
\be\label{bad}
\int_{\im\  z=0} d^4z\ W_2(z) f(z)
\ee
is not well-defined. To derive the right prescription, we recall that $f(z)$ is
an entire function, and replace the integral above by
\be\label{better}
\int_{\im\  z=\eta} d^4z\ W_2(z) f(z).
\ee
If $\eta^2>\ell^2$, the integral is well-defined. Deforming the integration
contour back to the real subspace, we find the prescription for treating the pole in
Eq.~(\ref{explicit}): 
\be
\frac{1}{z^2+\ell^2}\ra \lim_{\eps\ra 0}\frac{1}{z^2+\ell^2+i\eps z\eta}
\ee 
The latter distribution is Lorenz-invariant, but not invariant under $z\ra -z$.

There is a well-known theorem (the theorem about
global nature of local commutativity, see e.g.~\cite{BLT}) which, roughly speaking, says 
that if the fields (anti)commute for large enough space-like separations, then they 
(anti)commute for all space-like separations. This statement, when rephrased in
terms of Wightman functions, is clearly invalid in the above example.
The theorem about the global nature of local commutativity does apply because
one of the assumptions in its proof is that the Wightman functions are
tempered distributions.
 
\section{Approximately local observables in LSTs}

\subsection{General idea}

In the previous section we saw that operators in Little String Theories should
be smeared with entire test functions. Since a nontrivial entire function
cannot vanish in an open set, the corresponding observables are highly
nonlocal. On the other hand, at low energies LSTs flow to local quantum field theories
(Property 6). Thus the theories should be approximately local at distances
larger than ${\sqrt N}/M_s$. As explained in the Introduction, the microlocality
axiom is empty if only analytic test functions are allowed, so we need to find some
replacement for it which would ensure the locality of LSTs ``in the large.''

The weak quasilocality postulate discussed in the previous section ensures that 
Wightman functions are symmetric functions of their arguments when the arguments
are far apart and space-like separated. 
However, in order to be able to claim that a nonlocal theory flows to a local theory
in the infrared, one needs more than this. 
A local theory has local observables associated to compact sets. If a nonlocal
theory is approximately local in the infrared, there should be a way to define
approximately local (AL) observables  associated to compact sets which are ``large'' in 
some sense. Furthermore, one should require that AL observables approximately
commute (or anticommute) when their supports are space-like.

In the case of Little String Theories, we know for a fact that they flow to local
quantum field theories in the infrared, so understanding approximately local
observables is of paramount importance.

The problem of defining AL observables in nonlocal field theories
was previously addressed in~\cite{S,FSfirst,FS}. The present section is our interpretation
of~\cite{FS}. We assume that the reader is reasonably comfortable with
the notion of a topological vector space. Readers with low tolerance for math
may skip this section on first reading; such readers should be warned that
the discussion below touches on some important physics of LSTs.

To define observables approximately localized on a closed set $M\subset \RR^d$,
it is natural to consider a sequence of test functions $\{f_\nu\}$ which converges
to zero in the open set $\RR^d\backslash M$ in some sense. Then observables 
$\phi(f_\nu)$ should be regarded as approximately localized in $M$, the approximation 
getting better as $\nu$ increases. To define ``convergence in an open 
set,'' we need a topology $\tau(\cO)$ on the space of test functions for each open set 
$\cO\subset \RR^d$. The meaning of $\tau(\cO)$ is the following: two test functions are ``close''
in the topology $\tau(\cO)$ if and only if they are ``close'' everywhere in $\cO$.
We then say that a sequence of test functions is localized on a closed set $M$ if it is
convergent to zero in the topology $\tau(\RR^d\backslash M)$. We will also say that
$M$ is a quasisupport of the sequence $\{f_\nu\}$.

The main problem is how to choose the topologies $\tau(\cO)$.
A natural restriction on the choice of topologies is that if $\cO_1\subset \cO_2$,
then $\tau(\cO_1)$ should be weaker than $\tau(\cO_2)$. In other words, if
two functions are ``close'' on $\cO_2$, they should be ``close'' on $\cO_1$.
Another natural restriction is to require that $\tau(\RR^n)$ be the same as the
original topology on the space of test functions. Indeed, a sequence $\{f_\nu\}$
converging to zero in the topology $\tau(\RR^n)$ should be regarded as approximating a 
function which is identically zero. Then it is natural to require that $\phi(f_\nu)$
converge to zero. Thus all fields must be continuous functionals in the topology
$\tau(\RR^n)$. The original topology on $S_g$ has this property by definition, and
in general there is no other natural topology with this property.

What is the ``original'' topology on $S_g$? One way is to define it is to use
the family of norms~(\ref{norms}) to define convergence on $\tS_g$ and then apply
Fourier transform. We are going to use an equivalent definition~\cite{GelfShi,FS}
which makes use of the fact that all functions in $S_g$ can be analytically continued
to $\CC^d$. The topology on $S_g$ can be specified by saying which sequences of functions
converge to zero. We declare that a sequence $\{f_\nu\}$ converges to zero in the
topology $\tau(\RR^n)$, if it converges to zero uniformly in all sets of the form 
\be
V_a=\left\{x+iy\stl x,y\in \RR^d, ||\im\  y||\leq a\str\right\}, \quad a>0.
\ee
One can check that with this choice of topology $S_g$ becomes a complete 
countably normed Montel space~\cite{GelfShi}.

If we do not assume anything about the quasilocal theory in question, then the
only natural choice for $\tau(\cO)$ seems to be the topology of uniform convergence in 
all sets of the form 
\be
V_a(\cO)=\left\{x+iy\stl x\in \cO, y\in \RR^d, ||\im\  y||\leq a\str\right\}, \quad a>0.
\ee
This family of topologies satisfies both of the above requirements.

However, despite appearances, this choice of $\tau(\cO)$ does not really allow to 
define nontrivial observables associated to compact sets. Indeed,
consider a compact set $M$, and a sequence of functions converging to zero
in the topology $\tau(\RR^d\backslash M)$. According to our definition, we 
say that such a sequence of functions is localized on $M$. However, it turns out
that any such sequence actually converges uniformly to zero everywhere! (The proof
of this fact is very simple and is left as an exercise for the reader. 
See also~\cite{FS}, Section 1.8, where a similar statement is proven.) This means that 
if $M$ is any compact closed set, 
$\tau(\RR^d\backslash M)$ coincides with $\tau(\RR^d)$. Consequently, this family of
topologies does not allow to tell apart different compact sets of $\RR^d$, or even
to tell apart a compact set from the empty set.

The lesson here is that entire test functions are too smooth to allow a sensible
definition of quasisupport.

\subsection{Further constraints on the ultraviolet behavior of fields}

To do better than this, we need to impose some additional constraints on the 
high-energy behavior of fields. To motivate these constraints, we first define a 
new space of test functions $\tS^\ell_\eta$, where $\eta$ is a unit time-like vector and 
$\ell>0$ is a number. $\tS^\ell_\eta$ consists of all infinitely differentiable functions 
on $\RR^d$ all of whose derivatives decay faster than
\be
\exp(-\ell||p||_\eta).
\ee
Here $||p||_\eta^2=2(p\cdot \eta)^2-p^2$ is a Euclidean norm on $\RR^d$ associated
with the vector $\eta$.
The space $\tS^\ell_\eta$ was first introduced by Shilov~\cite{Shil} and studied in detail
in~\cite{GelfShi}. With a natural choice of topology $\tS^\ell_\eta$ becomes a complete 
countably normed Montel space, just like $\tS_g$.

Obviously, if $\ell>\ell'$, then $\tS^\ell_\eta\subset \tS^{\ell'}_\eta$. Thus for any 
fixed $\eta$, we have an infinite decreasing sequence of spaces
\be
\tS^1_\eta\supset \tS^2_\eta\supset \tS^3_\eta\supset\ldots
\ee
It is easy to see that our basic space of momentum-space test functions $\tS_g$ is the
intersection of all these spaces:
\be
\tS_g=\bigcap_{\ell=1,2,\ldots} \tS^\ell_\eta.
\ee
Moreover, one can check that the standard topology on $\tS_g$ is the direct 
limit of the standard topologies on $\tS^\ell_\eta$.

The Fourier transform of $\tS^\ell_\eta$ will be denoted $S^\ell_\eta$. 
According to \cite{GelfShi}, $S^\ell_\eta$ consists of functions which decay faster than
any polynomial at infinity and can be continued analytically into a strip
\be\label{strip}
\left\{x+iy|x,y\in \RR^d, ||\im\  y||_\eta <\ell\right\}.
\ee
The crucial difference between $S^\ell_\eta$ and $S_g$ is that in the latter case all
test functions are entire, while in the former case they can only be continued into
a strip of width $\ell$ off the real slice $\RR^d\subset\CC^d$.

Our basic space of test functions $S_g$
is the direct limit of spaces $S^\ell_\eta, \ell=1,2,\ldots$. The standard topology on
$S^\ell_\eta$ is the topology of uniform convergence in all sets of the form
\be
V_a^\eta=\left\{ x+iy\stl x,y \in \RR^d, ||\im\  y||_\eta\leq a \str \right\}, \quad
0<a<\ell.
\ee

The space $S^\ell_\eta$ is not Lorenz-invariant and so cannot be used as the basic
space of test functions in a QFT. Nevertheless these spaces play
an important role in QFTs based on the space $S_g$. Namely, one can show that any 
matrix element of $\phi(x)$ is a distribution on $S^\ell_\eta$ for some $\ell$ and $\eta$
~\cite{FI,GelfShi}. In general, $\ell$ depends on the states between which $\phi(x)$ is 
sandwiched.

In order to define approximately local observables, we will require that all 
fields be well-defined operator-valued distributions on $S^\ell_\eta$ for some {\it fixed} 
$\ell$ and $\eta$. Then by Lorenz-invariance all fields are well-defined distributions
on $S^\ell_\eta$ for all $\eta$.

\subsection{Definition and properties of approximately local observables}

With this additional requirement it becomes possible to define AL observables 
associated to all compact sets of size bigger than $\ell$.
More precisely, for any unit time-like vector $\eta$ we can define a family
of topologies $\tau_\eta(\cO)$ which satisfies all the requirements stated above. 
Thus possible definitions of localization are labeled by $\eta$. It is tempting to 
interpret $\eta$ as the 4-velocity of a reference frame. Then we have a different 
notion of approximate localization for different frames.

It is clear what the definition of topologies $\tau_\eta(\cO)$ should be. We simply
take $\tau_\eta(\cO)$ to be the topology of uniform convergence on all sets of the form
\be
V_a^\eta(\cO)=\left\{ x+iy\stl x\in \cO, y\in \RR^d, ||\im\  y||_\eta\leq a, 
\str \right\},\quad 0<a<\ell.
\ee
It is easy to see that both requirements on the family $\tau(\cO)$ stated above
are satisfied.

We need to check now that this new family of topologies allows to tell apart
different compact sets in $\RR^d$. This problem was addressed in~\cite{FS}.
These authors showed that the family $\tau_\eta(\cO)$ can tell apart a 
compact set $M$ from the empty set only if $M$ is big enough. For example, let 
$M$ be a ball 
\be
B_a=\left\{x\stl x\in \RR^d, ||x||_\eta\leq a\str\right\}.
\ee
If $a<\ell$, then it turns out that any sequence of functions localized on $M$ converges 
uniformly to zero everywhere on $\RR^d$. In other words, for $a<\ell$ 
$\tau(\RR^d\backslash M)$ coincides with $\tau(\RR^d)$, and there are no nontrivial 
observables localized on $M$~\cite{FS}.
On the other hand, if $a\geq \ell$, then the topologies $\tau(\RR^d\backslash M)$
and $\tau(\RR^d)$ are different, and there are nontrivial observables approximately
localized on $M$~\cite{FSfirst,FS}.

One can give a criterion which determines if the compact set $M$ is
``big enough.'' The key mathematical input is the notion of the domain of 
holomorphy~\cite{Hormander}.
We say that an open set $\Omega$ in $\CC^d$ is a domain of holomorphy if there exists
a function $f$ which is holomorphic in $\Omega$ and cannot be analytically continued
to a bigger open set. For any open set $X$ we define its domain of holomorphy
$\Omega_X$ as the smallest domain of holomorphy containing $X$. If $X$ is itself a domain
of holomorphy, then $\Omega_X=X$, otherwise $\Omega_X$ is strictly larger than $X$.
Now let $M$ be a compact set in $\RR^d$ and let us set 
\be
X=\left\{ x+iy\stl x\in \RR^d\backslash M, ||\im\  y||_\eta<\ell\str\right\}.
\ee
It may well happen that $\Omega_X$ is the whole strip~(\ref{strip}). In this case
$M$ is too small, in the sense that the topologies $\tau(\RR^d\backslash M)$ and
$\tau(\RR^d)$ are equivalent. On the other hand, if $\Omega_X$ does not contain the whole strip
~(\ref{strip}), then convergence in the topology $\tau(\RR^d\backslash M)$ does not
imply convergence in the topology $\tau(\RR^d)$, and nontrivial observables associated 
to $M$ exist~\cite{FSfirst,FS}.

\subsection{The strong quasilocality axiom}

We have now defined the notion of AL observables (different for different reference 
frames). As discussed in Section I, if a theory is approximately local
at long distances, AL observables must approximately commute if their quasisupports
are space-like separated. More precisely, let $M$ and $N$ be two closed sets
which are space-like separated, and let $\{f_\nu\}$ and $\{g_\nu\}$ be sequences
of test functions (from the space $S^\ell_\eta$) whose quasisupports are
$M$ and $N$ respectively. Then the strong quasilocality axiom states that
\be
[\phi(f_\nu),\phi(g_\nu)]_-\ra 0\ or\ [\phi(f_\nu),\phi(g_\nu)]_+\ra 0\quad 
as\ \nu\ra\infty. 
\ee
Nonlocal QFTs satisfying this axiom will be called ``strongly quasilocal'', or simply
``quasilocal.''  

Our conjecture about the universality class of LSTs can now be stated very concisely. 
We propose that LSTs are strongly quasilocal QFTs in six dimensions, with
$\ell$ of order ${\sqrt N}/M_s$. 

It is shown in~\cite{FS} that strong quasilocality implies weak quasilocality.
Furthermore, it can be shown that weakly quasilocal QFTs obey the CPT and 
spin-statistics theorems~\cite{FI}. This agrees with what we know about LSTs.

With additional assumptions, such as the existence of a
mass gap, one can also define a unitary S-matrix in quasilocal QFTs and prove
Froissart-type bounds on the cross-sections~\cite{FItwo,FS}. However, LSTs do 
not have a mass gap, and consequently the S-matrix is not well-defined in these theories.

\subsection{An example of a quasilocal QFT}
Let us give a simple example of a QFT which satisfies the strong quasilocality axiom.
Let $\phi(m,x)$ be a one-parameter family of scalar fields satisfying
the commutation relation
\be
[\phi(m,x),\phi(m',x')]=\delta(m-m')\Delta_m(x-x'),
\ee
where $\Delta_m(x-x')$ is the commutator function of a free scalar field of 
mass $m$. In other words, for any $m$ the field $\phi(m,x)$ is a free scalar 
field of mass $m$,
and these fields commute for different $m$'s.
We define
\be
\phi(x)=\int_0^\infty dm \sqrt{\sigma(m^2)} \phi(m,x),
\ee
where $\sigma(t)$ is given by Eq.~(\ref{sigma}). Let $\Omega$ be the common
vacuum for all $\phi(m,x)$. We define the Hilbert space of
our theory as a completion of the space spanned by the vectors
\be
\phi(f_1)\phi(f_2)\ldots\phi(f_n)\Omega,
\ee
where $f_i$ are suitable test functions (see below), and $n$ runs over all 
nonnegative integers.

Obviously, the field $\phi(x)$ is 
Gaussian: all its $n$-point function are expressible as products of its 2-point 
function $W_2(x-x')$ using Wick's theorem. 
$W_2(x-x')$ is given by the formulas (\ref{explicit}),(\ref{better}) and is
a distribution on $S_g$ with $g(t)=\exp(\sqrt t)$. Thus we can take $S_g$ 
with $g(t)=\exp(\sqrt t)$ as our space of test functions.

This Gaussian quasilocal theory is reminiscent of the toy model discussed in Appendix
C of~\cite{MinSei}. The toy model is a scalar field in $1+1$ dimensions
whose mass undergoes a jump at $x=0$ but is otherwise free. This theory can be reinterpreted
as a holographic dual of a certain quantum-mechanical boundary theory living at $x=0$.
The boundary theory has a single operator $O(t)$ whose 2-point function grows exponentially in the
energy representation, and whose higher-order connected correlators vanish.
Our Gaussian QFT can be regarded as a higher-dimensional version of this boundary theory.

The commutator 
\be
[\phi(x),\phi(y)]=W_2(x-y)-W_2(y-x)
\ee
in our theory does not vanish outside the light-cone, as explained in 
subsection~\ref{toy}.
Instead it has a contribution proportional to
$$\delta\left((x-y)^2+\ell^2\right)\ \sign(x_0-y_0).$$

It is easy to see that the 2-point momentum-space Wightman function~(\ref{twopoint})
is well-defined 
as a distribution on $\tS^\ell_\eta$ for any $\eta$. Since the theory is Gaussian, 
this immediately implies that the same is true for all higher-point functions. 
Hence the theory satisfies the strong quasilocality condition.

For an example of an interacting weakly quasilocal QFT, see~\cite{Efimov}.

\section{Discussion}

Hopefully, by relating Little String Theories to quasilocal quantum field theories,
we have clarified the nature of the former, especially the degree to which
observables in Little String Theories can be localized. We argued that
there are no strictly local observables in LSTs, but there are observables which can
be approximately localized (with arbitrary accuracy) to regions whose size is bigger
than ${\sqrt N}/M_s$. Thus LSTs have a fundamental length scale which sets a limit
on the resolution which can be achieved by measuring physical observables. 
Somewhat unexpectedly, the notion of localization depends
on the reference frame: observer in different reference frames have different
procedures for measuring approximately local observables. This is achieved
by introducing a different space of test functions for each reference frame.
 
We also suggested that LSTs obey the strong quasilocality axiom which ensures 
that the theory is approximately local at large distances. This axiom is
a surrogate for the usual microlocality axiom and preserves most of the
usual consequences of microlocality (CPT invariance and the spin-statistics
relation)~\cite{FI,FS}.

On general grounds, it appears very natural that Little String Theories do not
have truly local observables. After all, interacting critical string field theory 
also appears to violate
locality~\cite{LSU,LPSTU}. What is surprising (at least to the author) is that this 
nonlocality can be accommodated simply by choosing a space of test functions different
from the usual Schwartz space $\cS$. 

We saw that many known properties of Little String Theories fit into the
framework of quasilocal field theory with $\ell\sim {\sqrt N}/M_s$. One important feature 
which we have not explained is T-duality. T-duality is related to the behavior of 
Little String Theory at distances of order $1/M_s$. Presumably, only very special 
quasilocal field theories enjoy this property. At any rate, the fact that observables 
in quasilocal field theories cannot be localized to distances 
shorter than the fundamental scale suggests that T-duality is not impossible for field 
theories in this class.

In this paper we have focused on the kinematics of LSTs, but of course one would like to
understand their dynamics as well. Since the usual definition of LSTs based on decoupling 
in critical string theory is very implicit, this is a hard problem. Some
progress in this direction has been made in~\cite{AhBer,GKP,GK}. Our hope is that the 
ultraviolet behavior of LSTs is simple enough. If this is the case, LSTs
may provide some insight into $(2,0)$ and $(1,0)$ superconformal
field theories to which they flow in the infrared.

If our proposal is correct, then $(1,1)$ LSTs realize
the old idea that a nonrenormalizable quantum field theory may arise 
as the infrared limit of a quasilocal field theory. (This was the primary
motivation for the study of quasilocal theories in 60's and 70's.) 
Indeed, consider a maximally supersymmetric super-Yang-Mills theory in six dimensions
with a simply-laced gauge group.
Such a theory is nonrenormalizable, and there seems to be no local quantum field 
theory which flows to it in the infrared. However, it emerges as the infrared limit
of a certain quasilocal field theory, namely an LST with $(1,1)$ 
SUSY~\cite{Sei}. 

A more speculative proposal is to try to use quasilocal field theories to
model nonlocality arising in critical string theory. In particular, we have in mind
applications to the Hawking information loss paradox. One popular viewpoint is that
information escapes from the black hole with Hawking radiation, even though this 
apparently violates causality. One might suspect that large causality violations are 
related to a huge Lorenz boost of the stationary observer at infinity relative to the 
stationary observer at the stretched horizon. The boost may ``magnify'' nonlocal 
effects inherent in string theory (see~\cite{LPSTU} and references therein). There are 
some concrete calculations in string field theory supporting this scenario~\cite{LSU,LPSTU}.
If this scenario is correct, then stringy effects can be large even when all curvature 
invariants are small. Perhaps quasilocal field theories on a curved background could 
provide a useful effective description of such situations. It is certainly suggestive
that in quasilocal theories the notion of an approximately local observable only makes
sense relative to a particular reference frame.

\vskip 12pt
\noindent {\bf Acknowledgements: }
I would like to thank Ofer Aharony, Korkut Bardakci, and Nati Seiberg for 
useful comments. This work was supported by a DOE grant DE-FG02-90ER4054442.

\end{document}